\documentclass[a4paper,fleqn, 10pt]{cas-dc}
\usepackage[numbers]{natbib}

\usepackage{amsthm}
\usepackage{pbox}
\usepackage{enumitem} 
\usepackage{booktabs, multirow, bigstrut}
\usepackage[flushleft]{threeparttable} 
\usepackage{pifont} 
\usepackage[T1]{fontenc} 
\usepackage{rotating, graphicx} 
\usepackage{makecell} 
\newcolumntype{P}[1]{>{\centering\arraybackslash}p{#1}}
\newcolumntype{M}[1]{>{\centering\arraybackslash}m{#1}}
\usepackage{adjustbox}
\newlength\mylength
\usepackage{acronym}
\usepackage{graphics}
\usepackage{placeins}
\def\tsc#1{\csdef{#1}{\textsc{\lowercase{#1}}\xspace}}

\usepackage{setspace}   

\usepackage[figurename=Fig.]{caption}
\usepackage[labelsep=period]{caption}
\tsc{WGM}
\tsc{QE}
\tsc{EP}
\tsc{PMS}
\tsc{BEC}
\tsc{DE}
\usepackage{algorithm, algorithmic}
\usepackage{chngcntr, array}
\theoremstyle{plain}
\newtheorem{assumption}{Assumption}

\newcommand*{\field}[1]{\mathbb{#1}}

\usepackage{upgreek} 
\usepackage{mathtools} 
\usepackage{tcolorbox} 
\newtheorem{definition}{Definition}
\theoremstyle{plain}
\newtheorem{remark}{Remark}
\newtheorem{proposition}{Proposition}
\usepackage{subfig}
\usepackage{tikz} 
\newcommand*\circled[1]{\tikz[baseline=(char.base)]{
            \node[shape=circle,draw,inner sep=2pt] (char) {#1};}}

\usepackage{comment} 
\usepackage{nomencl}
\makenomenclature

\begin{document}
\let\WriteBookmarks\relax
\def\floatpagepagefraction{1}
\def\textpagefraction{.001}

\shortauthors{H.T. Reda et~al.} 

\title [mode = title]{Adversarial Models Towards Data Availability and Integrity of Distributed State Estimation for Industrial IoT-Based Smart Grid}                      



\author[1]{Tasew Reda, Haftu}





\author[1]{Mahmood, Abdun}
\credit{Resources, Data curation, Writing - review and editing, Formal analysis, Validation, Supervision}
\author[2]{Anwar, Adnan}
\author[1]{Chilamkurti, Naveen}
\address[1]{Department of Computer Science and IT, La Trobe University,  Plenty Rd, Bundoora, VIC 3086, Australia}
\address[2]{School of IT, Deakin University,  75 Pigdons Rd, Waurn Ponds, IC 3216, Australia}



\begin{abstract}
Security issue of distributed state estimation (DSE) is an important prospect for the rapidly growing smart grid ecosystem. Any coordinated cyberattack targeting the distributed system of state estimators can cause unrestrained estimation errors and can lead to a myriad of security risks, including failure of power system operation. This article explores the security threats of a smart grid arising from the exploitation of DSE vulnerabilities. To this aim, novel adversarial strategies based on two-stage data availability and integrity attacks are proposed towards a distributed industrial Internet of Things-based smart grid. The former's attack goal is to prevent boundary data exchange among distributed control centers, while the latter's attack goal is to inject a falsified data to cause local and global system unobservability. The proposed framework is evaluated on IEEE standard 14-bus system and benchmarked against the state-of-the-art research. Experimental results show that the proposed two-stage cyberattack results in an estimated error of approximately 34.74\% compared to an error of the order of $10^{-3}$ under normal operating conditions. 
\end{abstract}


\maketitle

\section{Introduction} \label{sec:introduction}
A smart grid \cite{reda2022comprehensive} is considered as a blend of modern communications infrastructures, a plethora of cyber-physical systems, intelligent sensing and metering entities, and modern energy management systems (EMS) based on the optimization of energy, network availability, demand, and son. With the emergence of Industry 4.0, industrial Internet of Things (IIoT), 5G/6G, and fog/edge/cloud computing, intelligent EMS solutions are becoming significant tools for smart grid implementation.

Centralized state estimation (SE) \cite{abur2004power}, \cite{liu2011false}, \cite{yi2021multi}, \cite{anwar2022measurement} is the most prominent scheme in the EMS of a smart grid. However, in view of the rapidly changing operational requirements of today's smart grid, the traditional centralised SE model may no longer be suitable for the control and operation of an increasingly distributed environment, taking into account the increased adoption of distributed energy resources \cite{worighi2019integrating}, distributed computing, and large scale deployment of IIoT devices in smart grid infrastructure. As an alternative solution, distributed SE (DSE) 
\cite{korres2010distributed}, \cite{kang2017distributed}, \cite{gomez2011taxonomy} has attracted significant attention in academia due to its several benefits, such as improved communication overhead, reduced computation burden, high scalability, and high robustness.

Compared with the traditional centralized SE model, DSE seems to be a more feasible component of the emerging smart energy management and security solutions. Yet, the DSE has its own limitations. For example, inter-area communication within the distributed state estimators can be subject to random gross errors, failure in communication link, and security threats. Thus, security aspects of the DSE need to be robust against communication noise, communication delays, link failures, and unexpected changes due to cyberattacks. In particular, coordinated cyberattacks may intend to infringe local estimators across the distributed IIoT networks of the power grid environment. Local zones within the distributed environment can be easily protected by their respective grid operators; however, boundary measurements among the local zones are sensitive and very critical. Consequently, if such tie-line parameters are exposed to cyberattacks, this will result in perturbation of critical system states of the distributed grid, and failure of the power system operation. Thus far, many articles have been reported on DSE. However, there has been very little work on the impact of cyberattacks towards the DSE. It is therefore of utmost importance to explore potential vulnerabilities towards the distributed power system and look for security solutions. To fill the research gap, this paper proposes adversarial models towards the DSE of smart grid. 

This article's major contributions include the following: \circled{1} The SE problem is formulated using a distributed optimization approach based on the alternating direction method of multipliers (ADMM). Here, the formulated DSE is compared to a benchmark using a centralised-based weighted least squares (WLS) SE technique under normal operating conditions. \circled{2} A data availability attack is proposed to investigate if the cyberattack can prevent boundary data exchange across a communication link and result in local unobservability of the distributed power grid. \circled{3} In addition, a two-stage cyberattack is proposed to investigate if the combined attack of data availability and integrity can destabilize both the local zone where the attack was launched and the entire network. \circled{4} Finally, the proposed framework is evaluated using sets of metrics such as local and global state estimated error, mean squared error (MSE), and estimation error using $l_2$-norm under various scenarios.
 
The rest of this paper is arranged as follows. First, background on security aspects of SE of the power system is presented in Section \ref{backG}. Then, Section \ref{reltd} reviews the existing literature on bad data analysis in DSE. Next, problem formulation of the DSE via ADMM algorithm is discussed in Section \ref{probF}. The subsequent section discusses proposed framework and adversarial models methodology. Furthermore, Section \ref{reslt} discusses performance evaluation of the proposed framework. Finally, the paper is concluded in Section  \ref{concln}. 
\section{Background} \label{backG}
SE \cite{abur2004power}, \cite{yi2021multi} has been well practised in the industry for the last couple of decades. It is an essential component of the smart grid EMS, and its main functions include estimation of power system states and detecting bad data and/or cyberattacks based on measurements received from communication devices on a regular basis. The measurement model is given by
\begin{math} 
\mathbf{y} = \mathbf{H}\mathbf{x} + \mathbf{w}
\end{math} 
for a DC power flow model and 
\begin{math} 
\mathbf{y} = \mathbf{h}(\mathbf{x}) + \mathbf{w}
\end{math}
for an AC power flow model \cite{abur2004power}. Here, $\mathbf{y} = (y_1, y_2,...,y_{\mathcal{M}})^T$ such that $y_\nu \in\field{R}^{\mathcal{M}\times1}$ refers to measurement vector, $\mathbf{H}$ or $\mathbf{h}()$ $\in\field{R}^{\mathcal{M}\times \mathcal{N}}$, commonly known as the Jacobian matrix, is a  function that relates $\mathbf{y}$ and power system states $\mathbf{x} = (x_1, x_2,...,x_\mathcal{N})^T$ such that $\mathbf{x} \in\field{R}^{\mathcal{N}\times1}$. And, $\mathbf{w} \in\field{R}^{\mathcal{M}\times1}$ is an additive noise, which is attributable to the communication channel and is usually modeled to follow additive white Gaussian noise (AWGN) distribution.

WLS \cite{abur2004power}, \cite{liu2011false}, \cite{korres2010distributed}, \cite{kang2017distributed} is one of the most widely used centralised SE method that uses minimization of weighted sum of the squares of residual errors. 
Although WLS has a good performance for the traditional centralised SEs, it may not be viable for IIoT-based smart grid characterised by distributed renewable energy sources, ubiquitous networked devices, and distributed computing environment. Meanwhile, with the advent of a distributed management system \cite{reda2022comprehensive}, the implementation of SE necessitates a more distributed solution. Moreover, a report has indicated that the WLS-based centralised SE can be evaded by false data injection attacks \cite{liu2011false}, \cite{yi2021multi}, \cite{anwar2022measurement}, \cite{reda2022data}. To overcome these limitations, DSE is eminently required for the ubiquitous IIoT-based smart grid. DSE has also been very promising in practice where the practical application of the DSE can be found in the multi-zone state estimators of smart grids, which may be run by separate system operators, each of which trying to determine its own system states \cite{gomez2011taxonomy}.
\section{Related Work}\label{reltd}
In recent past, a great deal of effort has been made in developing DSE techniques and the security issues related with them. The majority of DSE techniques may be classified as either hierarchical or fully distributed \cite{gomez2011taxonomy}. Hierarchical DSE approaches work on two-levels: first, a local estimation using local measurements across individual power system zones, followed by a final aggregated estimation using the local estimations and boundary information at a central coordinator. The authors of \cite{korres2010distributed} proposed hierarchical-based DSE and distributed bad data analysis for multi-zone power systems. For the bad data analysis,  each zone computes local residue vectors and gain matrices associated with the boundary measurements and sends the results to a central coordinator. The coordinator then exploits the communicated data to determine if any gross errors corrupt the measurements using largest normalised residue (LNR) \cite{abur2004power} test. A similar line of research in  \cite{kang2017distributed} proposes bad data detection in a distributed method of WLS-based SE in multi-zone power system. In \cite{kang2017distributed}, the proposed SE leverages weight updates of actual and pseudo measurements at both local and global levels to improve accuracy of the local and global state estimators. LNR test is used to assess whether or not the measurements involve  bad data. Similarly, the authors of \cite{guo2016hierarchical} proposed distributed bad data identification using LNR test based on the hierarchical DSE where the central coordinator exploits the sensitivities
of local objective functions with respect to boundary states. Although the related works (\cite{korres2010distributed}, \cite{kang2017distributed}, \cite{guo2016hierarchical}) introduced a robust DSE against bad data as compared to the centralized version, there are some drawbacks. Foremost is the hierarchical approach where estimations are performed at two levels: local zones and the central coordinator. Although the hierarchical distributed methods can alleviate communication overheads to a certain degree, as the power system grows in a very large scale, computational complexity and communication congestion issues can occur because of the presence of the centralized coordinator. Another drawback is that the bad data analysis is investigated at the central coordinator ignoring the impact of the bad data at the local SEs and global level. Moreover, the proposed bad data analysis techniques are based on LNR test, a conventional bad data detector which is found to be vulnerable to false data injection attacks \cite{reda2022comprehensive}, \cite{liu2011false}, \cite{anwar2022measurement}. 

\begin{table*}
\caption{Comparison of the proposed scheme with existing works.}
\label{relW} 
\begin{tabular}{|m{0.5cm}|m{0.7cm}|m{0.7cm}|m{2cm}|m{0.9cm}|m{1cm}|m{1.3cm}|m{1.3cm}|m{4.25cm}|} \hline 
& \multicolumn{8}{c|}{Comparative feature} \\ \cline{2-9}
 Lit.& SE & Power flow& Adversarial approach& Attack on integrity& Attack on availability &Adversarial impact (local) &Adversarial impact (global) & Evaluation metrics\\ \hline  \cite{korres2010distributed} & HDSE &DC & Gross error on local zone measurements & \ding{53}& \ding{53} & \ding{53} & \ding{53} &Computation time, measurement evaluation against normalised resideu test\\ \hline 
 \cite{kang2017distributed} & HDSE &AC & Gross error on boundary measurements & \ding{53} & \ding{53}  & \ding{53} & \ding{53} &State estimated error \\ \hline
  \cite{guo2016hierarchical}
& HDSE &DC& Gross error on local zone measurements & \ding{53} & \ding{53}  & $\checkmark$ & \ding{53} &Computation time and state estimated error \\ \hline 
 \cite{xie2012fully}& FDSE&DC and AC & \ding{53} & \ding{53} & \ding{53} & \ding{53} & \ding{53}&Computation time and estimation accuracy between proposed DSE and centralised SE \\ \hline 
 \cite{li2020fully}
 & FDSE&DC and AC& Gross error on meter measurements & \ding{53} & \ding{53} & $\checkmark$ & \ding{53} &MSE \\ \hline
\cite{zhang2016multiagent}& FDSE&DC & Bad data at local measurements & $\checkmark$ & \ding{53} & $\checkmark$ & \ding{53}&State estimated error \\ \hline
 \cite{yang2021distributed}
 & FDSE&AC& Bad data at local and boundary measurements & $\checkmark$ & \ding{53} & $\checkmark$ & \ding{53} &MSE \\ \hline 
\cite{kekatos2012distributed} & FDSE &DC& Data integrity attack at local zones & $\checkmark$ & \ding{53}  & $\checkmark$ & \ding{53} &Per-zone state estimated error (compared against ground truth and centralised WLS, MSE) \\ \hline
\cite{du2019admm}
 & FDSE&DC& Data deception and DoS attacks & $\checkmark$ & $\checkmark$ & $\checkmark$ & \ding{53} &State estimated error \\ \hline 
This paper & FDSE&AC & Two-stage data availability and distributed false injection attacks & $\checkmark$& $\checkmark$  & $\checkmark$ & $\checkmark$ &Per-zone and global state estimated error, per-zone and global $l_2$-norm estimation error, per-zone and global state estimated error under various scenarios, MSE, and attack magnitude.\\ \hline 
\end{tabular} 
\begin{tablenotes}
\item \hspace{30mm} HDSE: Hierarchical DSE; FDSE: Fully DSE; $\checkmark$: considered; \ding{53}: not considered
\end{tablenotes}
\end{table*}
 
Smart grid is turning into larger and more complex cyber-physical system as increasingly more IIoT network devices, and renewable and distributed energy resources are incorporated \cite{worighi2019integrating}. Consequently, the grid computing is shifting into a more fully distributed paradigm. Unlike to the hierarchical-based distributed state estimators, the fully DSEs \cite{xie2012fully}, \cite{li2020fully}, \cite{zhang2016multiagent}, \cite{yang2021distributed}, \cite{kekatos2012distributed}, \cite{du2019admm} do not require a coordinator and hence offer significantly better performance in terms of communication overhead and reduced computational burden. In fully distributed SE approaches, the global SE result is obtained through localized estimations and a minimal information exchange among the neighbors. In \cite{xie2012fully}, a fully DSE algorithm is proposed for wide-area monitoring in power systems. The authors showed that no coordinator is needed for each local zone to obtain provable estimation convergence of the integrated grid to that of the WLS-based centralized SE. Similarly, in \cite{li2020fully}, a bus-level DSE based on graph-theoretic and the conventional WLS algorithm is proposed. One of the potential advantages of \cite{li2020fully} is that the approach does no longer require partitioning of the power system as the SE is based on a bus-level; however, it incurs very high computational and communication costs, especially with a large number of buses. 

Under normal conditions, the aforementioned techniques perform well; nevertheless, their performance can suffer in the presence of malicious data attacks. A fully DSE solution and network topology identification is proposed in \cite{zhang2016multiagent} leveraging the WLS algorithm where bad data is considered at local zone measurements. Moreover, in \cite{yang2021distributed}, the authors studied fully DSE problem using Kalman filter for large-scale power systems in the presence of bad data at both local and boundary measurements. Further, \cite{kekatos2012distributed} and \cite{du2019admm} studied the impact of cyberattacks in the fully DSE scheme. The DSE framework in \cite{kekatos2012distributed} jointly conducts SE and analyses attack identification using the LNR test. The technique has a benefit of more robust estimation especially under cyberattacks. However, there are certain drawbacks in the work. One is that the cyberattack is assumed to remain local as the main intention is to avoid utilizing the bad data during the iterative estimation procedures. This implies that all the participating zones are trusted during the updates. However, the distributed optimization's periodic information exchange can be vulnerable to the adversarial attacks. For example, a malicious user who is aware of the underlying network topology, or a more intelligent and coordinated adversary can compromise the exchanged data. In addition, the work in \cite{du2019admm} suggested a DSE framework for a joint SE and attack detection using residue norms across individual local zones. It is the first work to theoretically demonstrate the performance of a DSE in the presence of a single attack and simultaneous cyberattacks (namely, deception and denial of service (DoS) attacks). 

Contrary to the aforementioned approaches, the proposed framework investigates the security issues of the DSE, primarily from the adversarial point of view. The approach of this research is distinctive in the following ways: 
\begin{itemize}
\item First, unlike the previous state-of-the-art works, the proposed method explores the effect of cyberattack against both local control centers and the integrated grid under a fully distributed environment formulated via ADMM algorithm.
\item  Moreover, a two-stage cyberattack against the distributed environment is examined by proposing orchestrated data availability attack and distributed false injection attack.
\item  Furthermore, the distributed environment is evaluated using a variety of evaluation metrics under three distinct scenarios (i.e., normal circumstance, availability attack against a local SE, and availability and false injection attack against the integrated grid). In all cases, the proposed method is compared against ground truth system states and the centralised WLS SE technique.
\end{itemize}
Table \ref{relW} summarizes the comparison of this work to the previously published articles. To the best of our knowledge, the research work being introduced in this article is unique in providing significant security issues of a smart grid by focusing on adversarial strategies towards the distributed state estimator of the EMS.

\section{Problem Formulation}\label{probF}
Recently, it has become so natural to look to paralleled and distributed computation algorithms as a mechanism for exploring large-scale statistical problems. Distributed optimization \cite{boyd2011distributed,muhlpfordt2021distributed,molzahn2017survey,9040671} techniques allow us to decompose an optimization problem into smaller, more manageable sub-problems that can be solved concurrently. These have gained significant attention in a variety of applications, such as machine learning \cite{boyd2011distributed}, signal processing \cite{9040671}, and power flow \cite{muhlpfordt2021distributed}. A review of numerous distributed optimization methods can be found in \cite{boyd2011distributed}, \cite{molzahn2017survey}.

ADMM \cite{boyd2011distributed,muhlpfordt2021distributed,molzahn2017survey} is a powerful algorithm, very suitable for distributed optimization. It integrates the quality of the augmented Lagrangian and the method of multipliers with the
distributed processing function of dual decomposition. The ADMM-based DSE offers several advantages. First, it has the advantage of decomposing large-scale problems into sub-problems, allowing for completely distributed or decentralized solutions. Here, the distributed control centers try to solve independent and paralleled sub-problems and form a final solution. Moreover, the distributed control centers that use ADMM often have to exchange small volumes of information with a subset of other distributed control centers, minimizing communication overhead of central coordination and communication as much as possible. The ADMM-based SE approach also has benefits in terms of robustness in the face of individual distributed control center link failure.  Besides, because the ADMM-based distributed optimization can perform parallel computations, it has the potential to outperform centralized SE algorithms in terms of computational efficiency, such as achieving fast convergence rate. This makes the ADMM as a viable solution to a distributed IIoT-based smart energy grid. In this paper, the SE problem is formulated via the ADMM algorithm (hereinafter referred to as the ADSE).

The following two generalised assumptions are introduced on the distributed optimization problem. Comments based on these assumptions have also been included. 
\begin{assumption}[System Partitioning] \label{as:1} The power system is partitioned into $Z$ number of zones, with $\kappa = \{1, 2, ..., Z\}$ being the zone index.\end{assumption}
\begin{assumption}[Separability] \label{as:2} Following  Assumption \ref{as:1}, the objective function is separable with respect to the decomposion of system variables into sub-vectors/sub-matrices.\end{assumption}
Assumption \ref{as:1} signifies that each zone of the power system knows its local 
information including local measurement values $\mathbf{y}_\kappa$, Jacobian information $\mathbf{H}_\kappa$, and state vectors $\mathbf{x}_\kappa$ where partitioning of the state vectors complies with the subsets of buses in each subsystem. Hence, the measurement model can be formulated by \begin{math}
 \mathbf{y}_\kappa = \mathbf{H}_\kappa\mathbf{x}_\kappa + \mathbf{w}_\kappa
\end{math} for the DC power flow model or by  (\ref{eqnLocMeas}) for the AC power flow model.
\begin{equation}
 \mathbf{y}_\kappa = \mathbf{h}_\kappa(\mathbf{x}_\kappa) + \mathbf{w}_\kappa
\label{eqnLocMeas}
\end{equation}
Assumption \ref{as:2} implies that the objective function $J(x)$ is separable such that the the optimization problem can lead to a decentralized algorithm \begin{math}
J(\mathbf{x}) = \sum_{\kappa=1}^K j_\kappa(\mathbf{x}_\kappa) = \frac{1}{2}\sum_{\kappa=1}^K ||\mathbf{y}_\kappa - \mathbf{H}_\kappa\mathbf{x}_\kappa||^2_2
\end{math}. Furthermore, this poses that the subsystems to jointly solve the distributed optimization problem
\begin{math}
     \hat{\mathbf{x}}_\kappa = \arg\min\limits_{(\mathbf{x_\kappa})} \sum_{\kappa=1}^K j_\kappa(\mathbf{x}_\kappa).
\end{math}

\begin{figure}
	\centerline{\includegraphics[width=80mm]{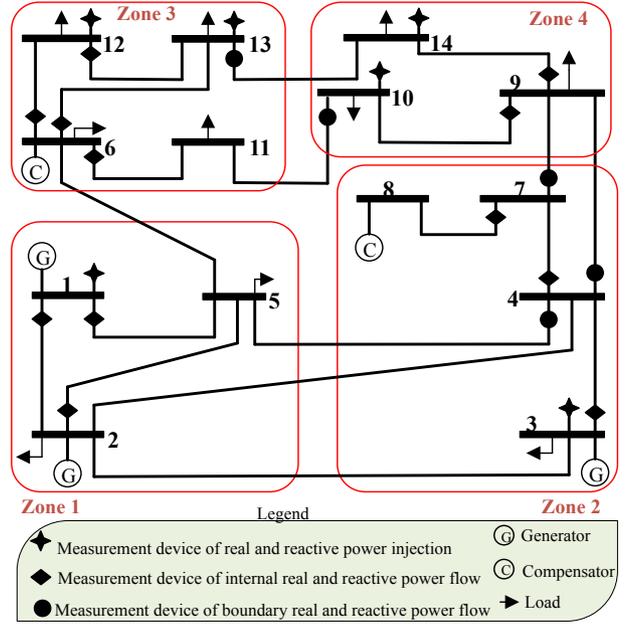}}
	\caption{Network topology of IEEE 14-bus system (with four distributed local zones), showing the bus interfacing and related measurement configuration.}
	\label{fig1}
\end{figure}
In the multi-zone system, a minimal information is communicated among the adjacent zones to account for the boundary measurement data \cite{korres2010distributed} \cite{kekatos2012distributed}. For example, see the boundary between $Z_1$ and $Z_2$ of Fig. \ref{fig1} where the associated sensors read power flow measurements across the transmission line between buses 5 and 4. Suppose $\upiota = \{1, 2, ..., K_\kappa\}$ is a neighbor of $\kappa = \{1, 2, ...,Z\}$ such that $\upiota \neq \kappa$. Let $\mathbf{x}_\kappa(\upiota)$ represents the subvector of $Z_\kappa$ shared with $Z_\upiota$. Similarly, let $\mathbf{x}_\upiota(\kappa)$ be the subvector of $Z_\upiota$ shared with $Z_\kappa$. Since the main objective of the distributed optimization problem is to enable each subsystem to independently solve \begin{math} \hat {\mathbf{x}}_\kappa = \arg\min\limits_{(\mathbf{x_\kappa})} \sum_{\kappa=1}^K j_\kappa(\mathbf{x}_\kappa)\end{math}, this imposes a constraint on the shared states, i.e. $\mathbf{x}_\kappa(\upiota) = \mathbf{x}_\upiota(\kappa)$. Further, let $\mathbf{x}_{\kappa\upiota}$ is the auxiliary  state variable such that $\mathbf{x}_\kappa(\upiota) = \mathbf{x}_{\kappa\upiota} = \mathbf{x}_\upiota(\kappa)$. Therefore, the distributed optimization problem can be posed as (\ref{Opt1}) subject to the constraint of (\ref{Opt2}).  
\begin{subequations}
\label{Opt}
\begin{align}
\hat{\mathbf{x}}_\kappa = \arg\min\limits_{(\mathbf{x_\kappa})} \sum_{\kappa=1}^K j_\kappa(\mathbf{x}_\kappa) \label{Opt1} \end{align}
 \begin{align}
 \text{s. t.}  \mathbf{x}_\kappa(\upiota) = \mathbf{x}_{\kappa\upiota} 
\label{Opt2}
 \end{align}
\label{Optt}
\end{subequations} 
Considering the objective function and constraints of (\ref{Opt}) and following the standard ADMM algorithm \cite{muhlpfordt2021distributed}, \cite{boyd2011distributed}, the Augmented Lagrangian function (\ref{eqnADMM1}) is considered for the multi-area power system 
\begin{equation}
\begin{split}
 L_\rho\bigg(\mathbf{x_\kappa},\mathbf{x}_{\kappa \upiota },\mathbf{\lambda}_{\kappa \upiota}\bigg) = \sum_{\kappa=1}^K \Bigg(j_\kappa(\mathbf{x}_\kappa) +  \sum_{\upiota=1}^{K_\kappa}\bigg(\mathbf{\lambda}_{\kappa \upiota }^\textit{T}\big(\mathbf{x}_\kappa(\upiota) -  \mathbf{x}_{\kappa\upiota}\big) + \\
 \bigg(\frac{\rho}{2}\bigg)\big|\big|\mathbf{x}_\kappa(\upiota) - \mathbf{x}_{\kappa\upiota}\big|\big|_2^2\bigg)\Bigg) 
\label{eqnADMM1}
\end{split}
\end{equation} 
where $\lambda_{\kappa \upiota}$ is the Lagrange multiplier, K$_\kappa$ refers to the set of neighboring nodes adjacent to the $\kappa^{th}$ zone, and $\rho>0$ is Lagrange penalty parameter. 

\begin{algorithm}
 \caption{Proposed ADSE}\label{admm}
 \begin{algorithmic}[1]
 \STATE \textbf{procedure} \textit{MAIN\_DISTRIBUTED\_SE\_PROGRAM}
 \STATE \: \textbf{Input}: Power flow parameters
  \STATE \: \text{Initialize} $\mathbf{x}_\kappa$, $\mathbf{H}_\kappa$, and measurement functions
   \STATE \: \text{Initialize} $maxIteration$
 \STATE \: $i \leftarrow 0$
 \STATE \: \textbf{while} $i \leq maxIteration$ \textbf{do}
 \STATE \: \: Compute measurement functions, $\forall \kappa \in \mathcal{S}$
 \STATE \: \: Compute Jacobian matrices, $\forall \kappa \in \mathcal{S}$
 \STATE \: \:  \textbf{procedure 1}: Each zone $Z_\kappa$ iteratively estimates its $\mathbf{x}_\kappa$ \begin{subequations}
\begin{align}
 \mathbf{x}_\kappa^{i+1}:= arg\min\limits_{(\mathbf{x}_\kappa)}  L_\rho(\mathbf{x}_\kappa,\mathbf{x}_{\kappa \upiota}^i,\mathbf{\lambda}_{\kappa \upiota}^i), \label{minX}
\end{align}
\STATE \: \: \textbf{procedure 2}: Compute boundary states 
\begin{align}
\mathbf{x}_{\kappa \upiota}^{i+1}:= arg\min\limits_{(\mathbf{x}_{\kappa \upiota})}  L_\rho(\mathbf{x}_\kappa^{i+1},\mathbf{x}_{\kappa \upiota},\mathbf{\lambda}_{\kappa \upiota}^i), \label{minXkl}
\end{align}
\STATE \: \:  \textbf{procedure 3}: Update Lagrange multipliers 
\begin{align}
\mathbf{\lambda}_{\kappa \upiota}^{i+1}:= \mathbf{\lambda}_{\kappa \upiota}^i + \rho(\mathbf{x}_\kappa(\upiota)^{i+1} - \mathbf{x}_{\kappa\upiota}^{i+1}), \label{minLamdakl}
\end{align}
\end{subequations}
 \STATE \: \textbf{end while}
\STATE \textbf{end procedure}
 \end{algorithmic} 
 \end{algorithm} 
 
The proposed ADSE involves three main procedures as shown in Algorithm \ref{admm}. The algorithm begins by defining power flow parameters including bus/branch data, and observed measurement values. Next, subvectors of each zone are initialized, namely $\mathbf{x}^0_\kappa$, $\mathbf{H}^0_\kappa$, $[\mathbf{h}_\kappa(\mathbf{x})_\kappa]^0$ $\forall \kappa$. After that, at each iteration $i \geq 0$ the vectors $\mathbf{x}_\kappa$, $\mathbf{x}_{\kappa\upiota}$, and $\mathbf{\lambda}_{\kappa\upiota}$ are updated using  (\ref{minX}), (\ref{minXkl}), and (\ref{minLamdakl}), respectively. (\ref{minX}) refers to the minimization of the local state vectors and is performed independently across each zone, in distributed and parallel way. (\ref{minXkl}) and (\ref{minLamdakl}), respectively refer to the minimization of the shared states and the update of Lagrange multiplier. Following the approach in \cite{kekatos2012distributed} the closed-form solution to (\ref{minX}) is given by (\ref{stateADMM})  
\begin{equation}
\mathbf{x}_\kappa^{i+1}:=
 \bigg(\big(\mathbf{H}_\kappa^T\mathbf{D}_\kappa\mathbf{H}_\kappa\big) + \rho\mathbf{C}_\kappa\bigg)^{-1} \bigg(\mathbf{H}_\kappa^T\mathbf{y}_\kappa + \rho\mathbf{C}_\kappa \mathbf{q}_\kappa^i\bigg) 
\label{stateADMM}
\end{equation}
where the updates $\mathbf{s}_\kappa$ and $\mathbf{q}_\kappa$ are respectively given by 
\begin{math}
\mathbf{s}_\kappa^{i+1}:= \frac{1}{K_\kappa}
\sum_{\upiota \in K_\kappa}\mathbf{x}_\upiota^{{i+1}}, \end{math}
and \begin{math}
\mathbf{q}_\kappa^{i+1}:= \mathbf{q}_\kappa^i + \mathbf{s}_\kappa^{i+1} - 0.5(\mathbf{x}_\kappa^i + \mathbf{s}_\kappa^i) \label{qk1}.
\end{math}
And, $\mathbf{C}$ is a diagonal matrix such that  $\mathbf{C} \in \field{R}^{K_\kappa \times K_\kappa}  $.

\section{Methodology} \label{methdlg}
\subsection{System Model}
\begin{figure}
	\centerline{\includegraphics[width=80mm]{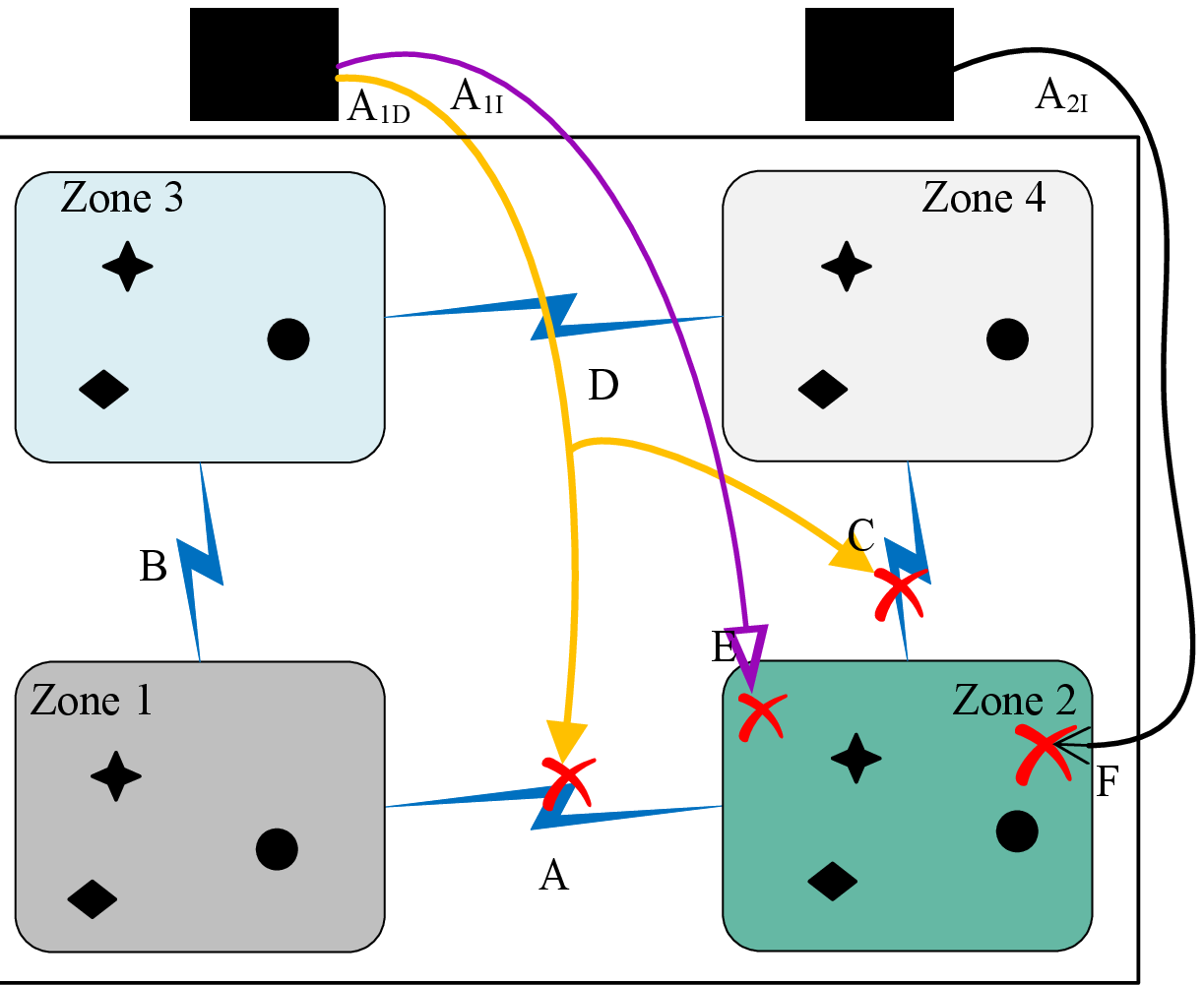}}
	\caption{Boundary data exchange under adversarial attacks.}
	\label{comm} {\footnotesize In Fig. \ref{comm}, $A_{1D}$ is a data availability attack against boundary data across the communication links. On top of that, a data integrity attack ($A_{1I}$) against Zone 2 is considered. The attack type $A_{2I}$ is a data integrity attack whilst the boundary data exchange is happening among the neighbors. While $A_{1D}$ and $A_{1I}$ happen one event after the other, the $A_{2I}$ attack type is mutually exclusive of the former.\par}
\end{figure}

Consider a wide-area monitoring of smart grid communication comprising of IIoT sensor networks. Fig. \ref{fig1} shows single-line  diagram of the standard IEEE 14-bus system. The network topology is partitioned in to four non-overlapping zones, namely $Z_1$, $Z_2$, $Z_3$, and $Z_4$, similar to the power system partitioning adopted in \cite{korres2010distributed}. Each zone has internal buses, internal branches, boundary buses and boundary branches.
Each local SE obtains its own as well as the entire grid system states by leveraging the internal and boundary data. For the AC model, four different types of power system quantities are considered, namely, real power injection, reactive power injection, real power flow, and reactive power flow. Bus interfacing, configurations, types and locations of measurements are shown in Fig. \ref{fig1} and Table \ref{sysMod}. Fig. \ref{comm} shows the boundary data exchange under adversarial attacks. In Fig. \ref{comm}, the communication links for boundary data exchange among the zones are labeled as $A$, $B$, $C$, and $D$.

\subsection{Proposed Cyberattack Framework}
Here, the proposed threat model is analysed towards the ADSE formulated by Algorithm \ref{admm}. Our main intuition is to explore the impact of cyberattack on the DSE, specifically with the following two main attack goals:
\begin{enumerate}
 \item \textbf{Attack Goal 1 (AG1)}: To investigate if the cyberattack launched in a particular zone results in local zone unobservability. 
 \item \textbf{Attack Goal 2 (AG2)}: To investigate if the cyberattack launched in a particular zone results in a global unobservability.  
\end{enumerate}
\theoremstyle{definition}
In Fig. \ref{comm}, attack types $A_{1D}$ and $A_{1I}$ represent data availability and integrity attacks, respectively, and they are part of the AG1. Further, the integrity attack under the AG2 is represented by $A_{2I}$. Under the first scenario, a data availability attack can occur at points $A$, $B$, $C$, or $D$. However, it can be noted that the attack does not have to compromise all the communication links. Zone 2 ($Z_2$) is taken as a case study and disrupt links $A$ and $C$. In this case, data availability attack across the two links will isolate $Z_2$ and now it can be observed how it affects the distributed DSE of $Z_2$. Next, it is shown how the attacker can further destabilise $Z_2$ by injecting false data into $Z_2$. Finally, it is demonstrated how the false data attack injected into $Z_2$ can actually destabilise not only $Z_2$ but also the entire grid when the communication of the distributed system happens properly assuming that there is no data availability attack. This attack is represented by $A_{2I}$ under the AG2 scenario.
\begin{definition}[Local zone unobservability]
An attack vector $\mathbf{a}_\kappa$ is said to cause system local zone unobservability if, by removing the compromised sensors, results in non-optimal system states of the bus(es) adjoined to the sensors of that particular zone.\label{def1} \end{definition} 
\begin{remark}
For Definition \ref{def1}, two distinct assumptions are made. One assumption is that the communication medium of the distributed system is vulnerable to data availability attacks causing loss of boundary update. And, the other assumption is a data integrity attack against a local zone. To this end, we propose a two-stage data availability and integrity attacks towards the distributed smart grid.
\end{remark}

\begin{table*}[ht]
\caption{Measurement data meters configuration of the distributed grid.}
\label{sysMod}
\begin{tabular}{|m{0.65cm} |m{2.5cm}m{4.25cm}|m{2.5cm}m{4cm}|} \hline 
 & \multicolumn{2}{c|}{\underline{Internal}}& \multicolumn{2}{c|}{\underline{Boundary}} \\ 
Zone & Power injection & Power flow & Power injection & Power flow \\ \hline
 $Z_1$  & \{$M_1$\} & \{$M_{1-2}$, $M_{1-5}$, $M_{2-5}$\} & $-$ & $-$ \\ \hline
$Z_2$ &\{$M_3$\} & \{$M_{3-4}$, $M_{4-7}$, $M_{7-8}$\} & $-$ & \{$M_{4-5}$, $M_{4-9}$, $M_{7-9}$\}  \\ \hline
$Z_3$ & \{$M_{12}$\} & \{$M_{6-11}$, $M_{6-12}$, $M_{6-13}$, $M_{12-13}$\} & \{$M_{13}$\} & \{$M_{13-14}$\} \\ \hline
$Z_4$  & - & \{$M_{9-10}$, $M_{9-14}$\} & \{$M_{10}$, $M_{14}$\} & \{$M_{10-11}$\} \\ \hline  \end{tabular}  
\end{table*} 
\subsubsection{Data Availability Attack}
Suppose that the inter-zone communication is subject to a data availability attack considering a scenario when an attacker targets against availability of the boundary information. DoS \cite{tsiatsis2018internet}, \cite{5473884} attacks are the most common availability attacks against communication systems that can be constructed at the physical layer (e.g. jamming attack), at the data link layer (e.g. collision attack), at the transport layer (e.g. flooding attack), targeting routing protocols or in the form of replay attacks into the network. DoS jamming attacks \cite{5473884} are common towards shared networks where a malicious user can continually send unwanted signals to block any legitimate access to the communication medium. This work proposes a DoS attack that can prevent the individual zones from further sharing of their periodic boundary data. 
\begin{proposition}
Given Algorithm \ref{admm}, let $\tau$ be the maximum number of iterations during which the local estimators try to update their boundary information. The exchange of boundary data under the DoS attack in the distributed environment can be modeled as a binary state of birth-death process with the birth rate probability denoted by $p_{u, \kappa}^i$, which refers to the probability of occurrence of the boundary update and the death rate probability denoted by $p_{A_{1D}, \kappa}^i$, which refers to the probability of occurrence of the DoS attack.
 \label{prop1}
\end{proposition}
\begin{proof}
The events of boundary data transmission in the distributed network is considered for the proof. Without loss of generality, any successful transmission can be subject to a communication link failure and/or DoS attack, where the latter two cases lead to a drop of the data. So, assuming no communication link failure, the packet loss can only be attributable to the DoS attack. The birth and death rate probabilities can have values $p_{u, \kappa}^i \in \{0,1\}$ and $p_{A_{1D}, \kappa}^i \in \{0,1\}$, where
\begin{math}
p_{u, \kappa}^i =
\begin{cases}
    1, & \text{if boundary update transmitted at iteration $i$} \\
    0, & \text{otherwise}.
  \end{cases}
\end{math} And, \begin{math}
p_{A_{1D}, \kappa}^i =
\begin{cases}
    1, & \text{if DoS attack occurs at iteration $i$} \\
    0, & \text{otherwise}. \end{cases} \end{math} 
Thus, under these scenarios the exchange of boundary data can be modeled by the two random variables $p_{u, \kappa}^i$ and $p_{A_{1D}, \kappa}^i$.  
\end{proof}
Furthermore, assume that the probability of a boundary data availability being lost under the attack is denoted by $\zeta$, then the conditional probability of the update data that can be received by the local zones is given by (\ref{rx1}). 
\begin{equation}
  \pi_{r, \kappa}^i = p_{u, \kappa}^i p_{A_{1D}, \kappa}^i (1-\zeta) + p_{u, \kappa}^i (1-p_{A_{1D}, \kappa}^i)
  \label{rx1}
 \end{equation}
Under this condition, the update rule in (\ref{stateADMM}) becomes (\ref{qk2}).
\begin{equation}
\mathbf{q}_\kappa^{i+1}:= \bigg(\mathbf{q}_\kappa^i + \mathbf{s}_\kappa^{i+1} - 0.5(\mathbf{x}_\kappa^i + \mathbf{s}_\kappa^i)\bigg) \bigg(\pi_{r, \kappa}^i\bigg) 
\label{qk2}
\end{equation} 
There are two cases worth considering: when $\zeta =1$ and following the birth-death process of Proposition \ref{prop1} if either $p_{u, \kappa}^i = 0$ or $p_{A_{1D}, \kappa}^i = 1$. This yields $\mathbf{q}_\kappa^{i+1} = 0$ which indicates under a successful data availability attack the local zones will not receive any boundary update. In this case, the local zones need to retain the previous recent update (i.e. $\mathbf{q}_\kappa^{i+1} \leftarrow \mathbf{q}_\kappa^{i}$, $\forall i \leq \tau$). Consequently, when the distributed system is subject to the availability attack, then (\ref{stateADMM}) is reduced to (\ref{attAG11}) 
\begin{equation}
\mathbf{x}_\kappa^{i}:=
 \bigg(\big(\mathbf{H}_\kappa^T\mathbf{D}_\kappa\mathbf{H}_\kappa\big) + \rho\mathbf{C}_\kappa\bigg)^{-1} \bigg(\mathbf{H}_\kappa^T\mathbf{y}_\kappa + \rho\mathbf{C}_\kappa \mathbf{q}_\kappa^{i^*}\bigg) 
\label{attAG11}
\end{equation} where $i^* \leq i$ refers to an iteration index at which the last boundary update occurs just before the availability attack has occurred. 
\subsubsection{Data Integrity Attack} \label{inteAttk}
After a successful launch of the availability attack, a malicious injection attack $\mathbf{a}_\kappa$ is considered that can compromise measurement readings of a single zone. This attack is formulated in \ref{threatMod}. Now under this attack scenario, the SE solution becomes
\begin{equation}
\mathbf{x}_{\kappa, f}:=
 \bigg(\big(\mathbf{H}_\kappa^T\mathbf{D}_\kappa\mathbf{H}_\kappa\big) + \rho\mathbf{C}_\kappa\bigg)^{-1} \bigg(\mathbf{H}_\kappa^T\mathbf{y}_{\kappa, f} + \rho\mathbf{C}_\kappa \mathbf{q}_\kappa\bigg) 
\label{attAG12}
\end{equation} where $\mathbf{x}_{\kappa, f}$ and $ \mathbf{y}_{\kappa, f}$ refer to the attacked local states and compromised measurement values ($\mathbf{y}_\kappa + \mathbf{a}_\kappa$), respectively. 
\theoremstyle{definition}
\begin{definition}[Global unobservability]
The attack vector $\mathbf{a}_\kappa$ is said to cause global unobservability if, by removing the compromised sensors, results in non-optimal system states of the bus(es) adjoined to the sensors of multiple zones. 
\label{def2}
\end{definition} 
\begin{remark}
The cyberattack in Definition \ref{def2} occurs when there is an inter-zone information exchange. \end{remark} Under this scenario, the iterative version of (\ref{stateADMM}) and the update terms for the boundary exchange are used:
\begin{equation} 
\mathbf{x}_{\kappa, f}^{i+1}:=
 ((\mathbf{H}_\kappa^T\mathbf{D}_\kappa\mathbf{H}_\kappa\big) + \rho\mathbf{C}_\kappa)^{-1} (\mathbf{H}_\kappa^T\mathbf{y}_{\kappa, f} + \rho\mathbf{C}_\kappa \mathbf{q}_\kappa^i) 
\label{stateADMM31} 
\end{equation}
\begin{equation}
\mathbf{s}_{\kappa, f}^{i+1}:= \frac{1}{K_\kappa}
\sum_{\upiota \in K_\kappa}\mathbf{x}_{\upiota, f}^{{i+1}}
\label{stateADMM32}, 
\end{equation}
\begin{equation}
\mathbf{q}_{\kappa, f}^{i+1}:= \mathbf{q}_{\kappa, f}^i + \mathbf{s}_{\kappa, f}^{i+1} - 0.5(\mathbf{x}_{\kappa, f}^i + \mathbf{s}_{\kappa, f}^i) \label{stateADMM33}.
\end{equation} 

\begin{algorithm}
 \caption{Adversary model under the ADSE approach}\label{random}
 \begin{algorithmic}[2]
 \STATE \textbf{Input:} $\mu_\kappa$, $\textbf{H}_\kappa$, $\textbf{y}_\kappa$
 \STATE \textbf{Output:} \textbf{y}$_{\kappa_{false}}$
 \STATE \textbf{procedure} $CONSTRUCTATTACK$($\mu_\kappa$, $\textbf{H}_\kappa$, $\textbf{y}_\kappa$)
 \STATE \: $I_{\mu_\kappa}$ $\leftarrow$ \begin{math}
  randi([1, m], 1,\mu_\kappa);  \end{math}
 \STATE \: \textbf{if} {$i \in I_{\mu_\kappa}$} \textbf{then}
 \STATE \: \: $\mathbf{a}_{\mu_\kappa}$(i) $\leftarrow$  $\mathbf{H}_\kappa$ $\times$ $randn(\mu_\kappa,1)$;
 \STATE \: \: $\mathbf{y}_{\kappa_{false}}$(i) $\leftarrow$ $\mathbf{y}_\kappa$(i) + $\mathbf{a}_{\mu_\kappa}$(i);
 \STATE \: \textbf{end if} 
 \STATE \: \textbf{if} {$i \notin I_{\mu_\kappa}$} \textbf{then}
 \STATE \: \: $\mathbf{a}_{\mu_\kappa}$(i) $\leftarrow$ $zeros(\mu_\kappa,1)$
 \STATE \: \textbf{end if}
 \STATE \textbf{end procedure}
 \end{algorithmic} 
 \end{algorithm}
 
\subsection{Threat Model of the Data Integrity Attack}\label{threatMod} 
Consider a false data injection attack \cite{liu2011false}, \cite{reda2022data} vector of the form $\mathbf{a}_\kappa = \{a_{\kappa,m}\} \neq 0$ which can compromise a limited number of measurements. The malicious data attack can target a subset of sensors denoted by $\mu_\kappa$ adjoined to a particular node of the local estimator and cause system unobservability of the power grid. Note that the adversary cunningly selects sufficient number of $\mu_\kappa$ such that the resulting attack vector is unobservable which may ultimately result in the perturbation of the critical system states. Let $\mathcal{M}_\kappa$ be the total number of measurement/sensor readings across zone $Z_\kappa$. We say that the attack vector causes local unobservability if the compromised  $\mu_\kappa$ sensors (where $\mu_\kappa\subseteq\mathcal{M}_\kappa$) are removed and if the convergence of the corresponding states is not achieved within an acceptable estimation error. One of the possible adversarial approach can be to counterfeit $\mu_\kappa$ out of the $\mathcal{M}_\kappa$ measurements provided that the attacker has access to the $\mu_\kappa$ sensor readings.  
\begin{proposition}
Suppose that $\mathcal{I}_{\mathcal{M}_\kappa} = \{i_1, i_2,...,i_{\mu_\kappa}\}$ refers to the set of indices of the compromised sensors, the attack vector can be formulated through (\ref{randm})
\begin{equation}
\label{randm}
\mathbf{a}_\kappa(i_\nu)=
\begin{cases}
    \mathbf{H}_\kappa\mathbf{b}_\kappa , & \text{if $i_\nu$} \in I_{\mathcal{M}_\kappa} \\
    \mathbf{0}, & \text{if $i_\nu$} \notin I_{\mathcal{M}_\kappa}
  \end{cases}
\end{equation} 
where $\mathbf{b}_\kappa$ is a non-zero injection vector whose array size corresponds to the non-zero column size of $\mathbf{H}_\kappa$.
\label{prop2}
\end{proposition}
\begin{remark}
Equation (\ref{randm}) is a distributed adversarial model with the assumption that the malicious user is aware of network structure of the power system.
\end{remark}
The data integrity attack of Proposition \ref{prop2} is applicable for the two attack goals (i.e. AG1 and AG2). The attack construction procedure of (\ref{randm}) is presented in Algorithm \ref{random}.
    
\section{Performance Evaluation and Discussion} \label{reslt}
This section provides justification through numerical simulations and discussions of the results. 
\subsection{Test System Scenario}
To demonstrate the performance of the proposed framework, sets of experiments are conducted on the multi-zone scheme of the IEEE 14-bus system given in Fig. \ref{fig1}. For all numerical results of this work, AC power flow model with voltage magnitudes and angles are considered. A total of 46 measurement quantities are used for the entire grid (i.e., 8 in $Z_1$, 14 in $Z_2$, 14 in $Z_3$, and 10 in $Z_4$). Following the network partitioning of Fig. \ref{fig1}, the measurement readings are grouped into internal and boundary (see Table \ref{sysMod} for the different measurement configurations of each zone). Both the internal and boundary data include active and reactive power injection (i.e. at buses), and active and reactive power flow (i.e. along branches) of the power grid. The set of power injection measurements are indicated by \{$M_n$\} where $M_n$ refers to the deployed sensor at the $n^{th}$ bus. Similarly, the set of power flow measurements are indicated by \{$M_{n_1-n_2}$\} where $M_{n_1-n_2}$ refers to the deployed sensor along the branch or transmission line $n_1-n_2$. To make the measurements more realistic, the communication channel is modeled as AWGN noise where a mean $\mu_w = 0$ and variance $\sigma_w^2$ = $10^{-4}$ is added to each measurement quantity.

\begin{figure}[h!] \centering
\subfloat[]{\includegraphics[width=1\linewidth]{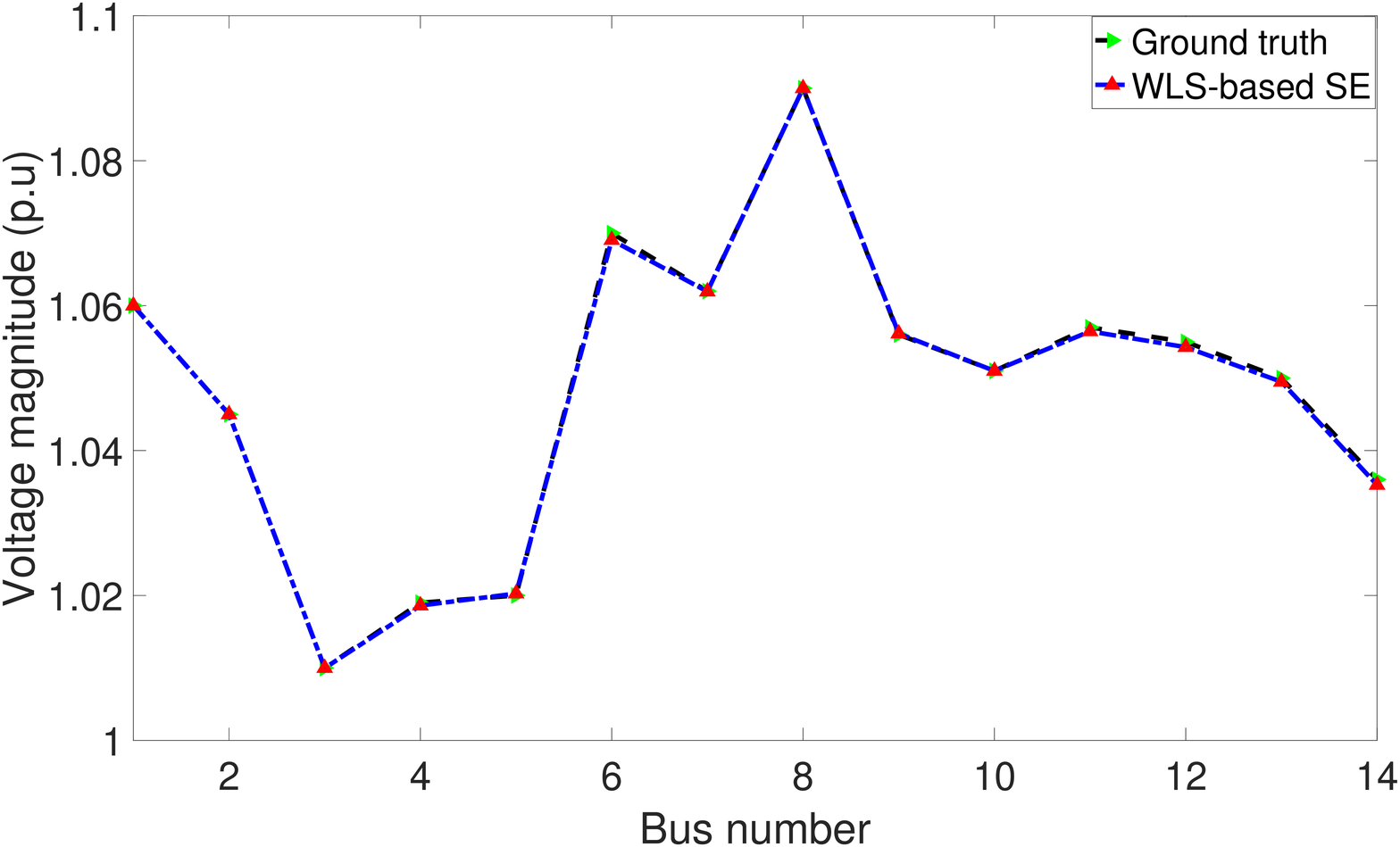}} \\
\subfloat[]{\includegraphics[width=1\linewidth]{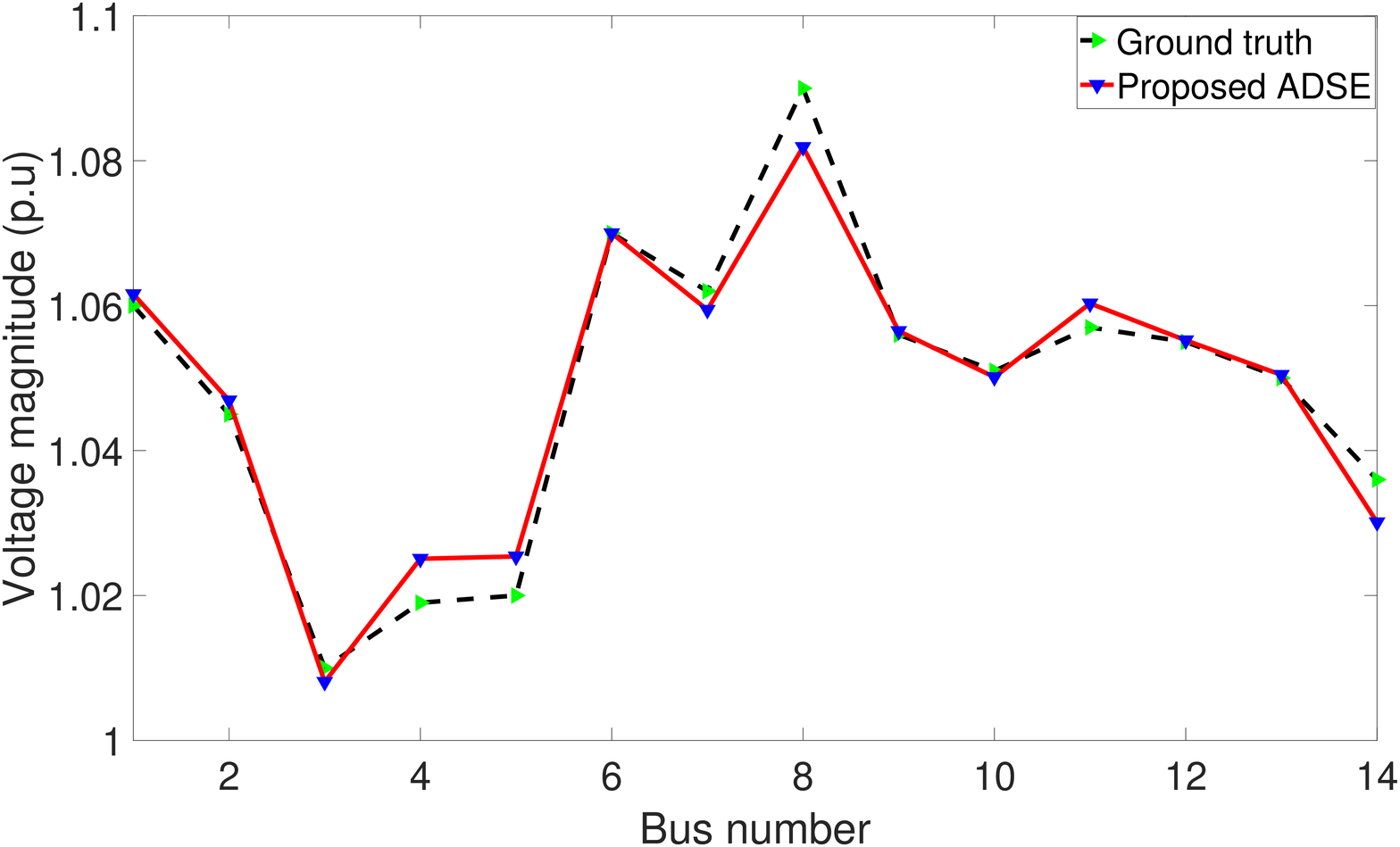}} \\
\subfloat[]{\includegraphics[width=1\linewidth]{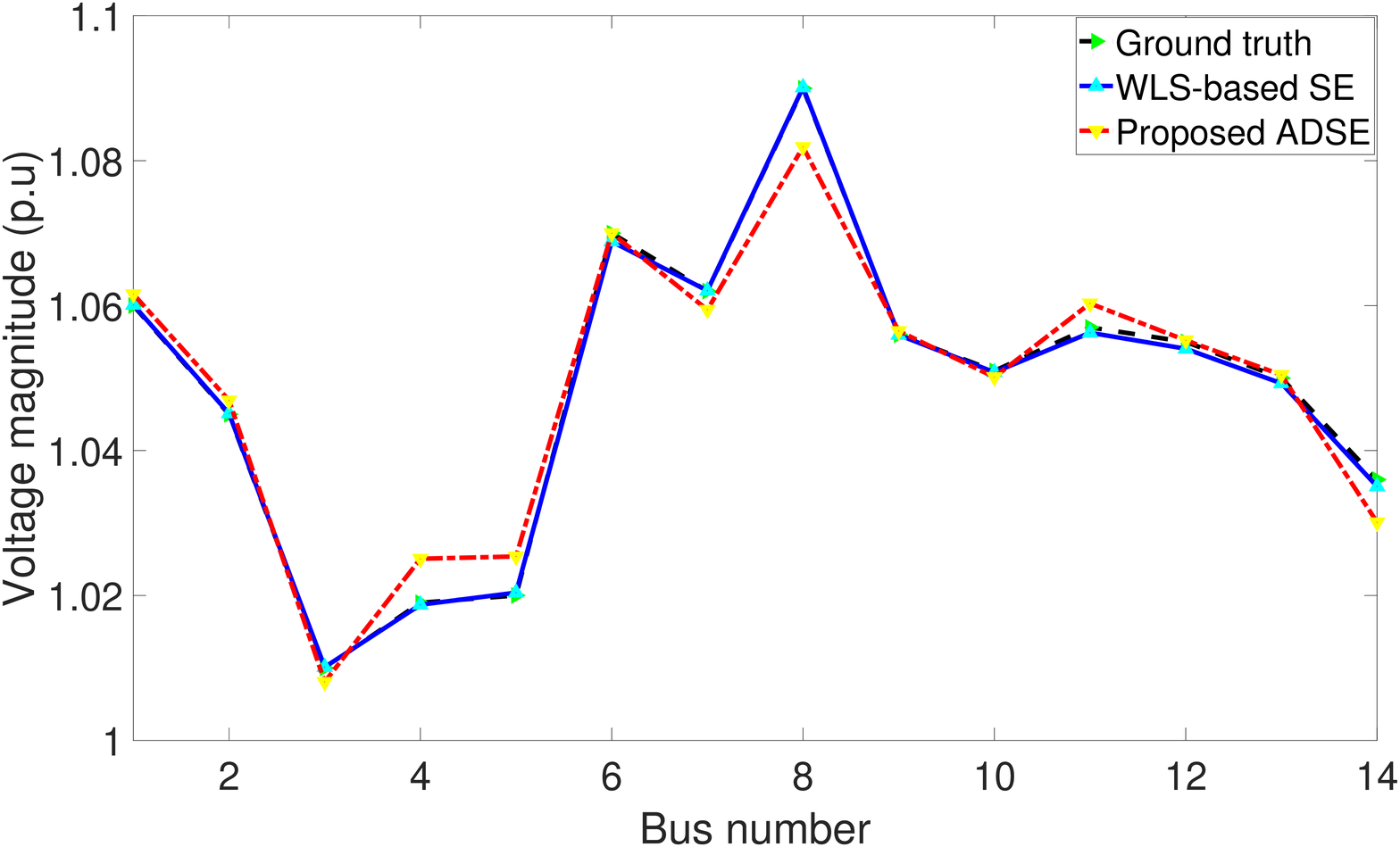}}
\caption{Estimation performance comparisons (considering $\sigma_w^2$ = $10^{-4}$) (a) Centralised SE against ground truth states, (b) ADSE against ground truth states, and (c) ADSE against ground truth states and the WLS-based centralised SE.} \label{est1}
\end{figure}

\begin{figure}[h!] \centering
\subfloat[]{\includegraphics[width=1\linewidth]{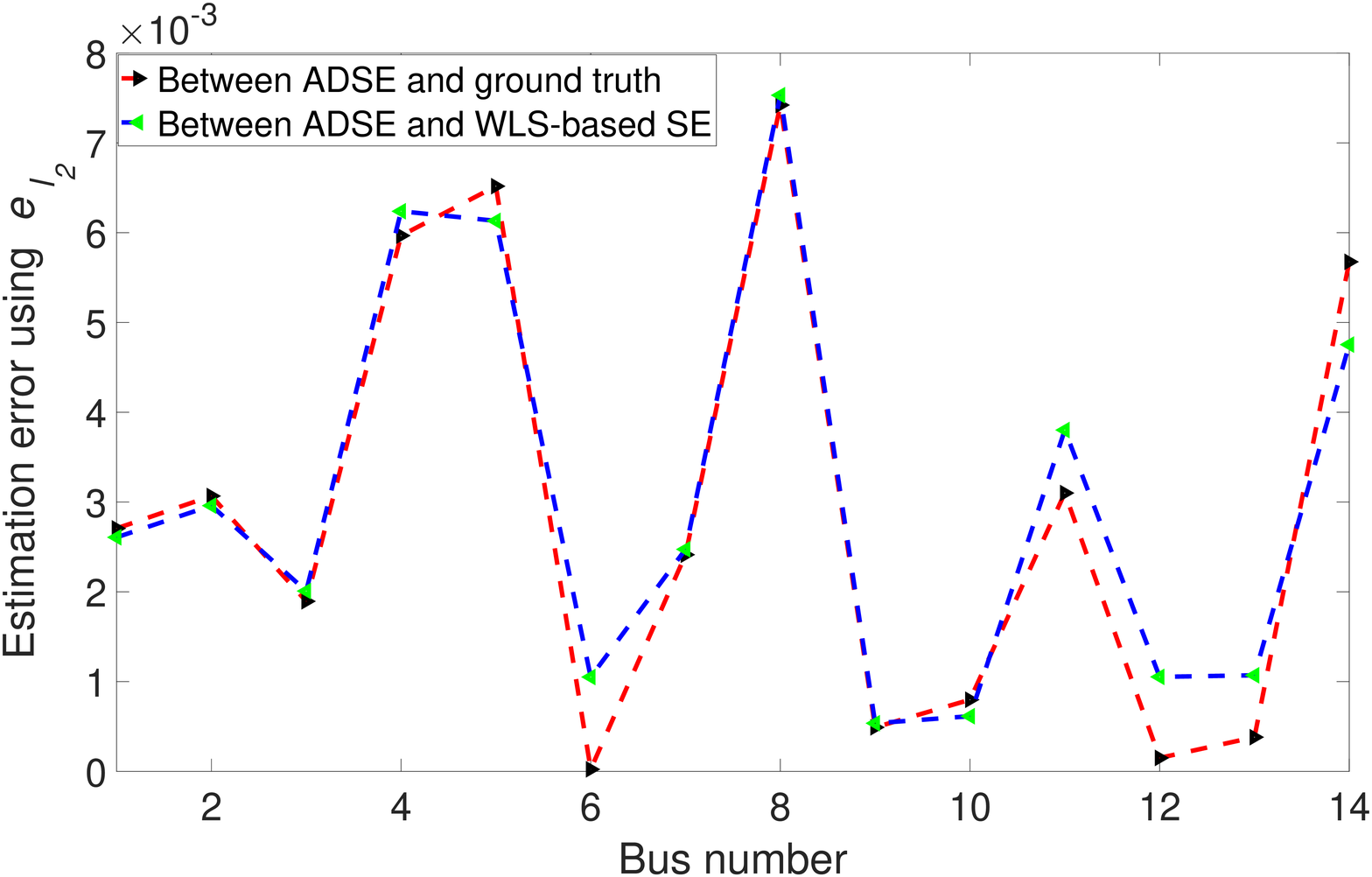}} \\
\subfloat[]{\includegraphics[width=1\linewidth]{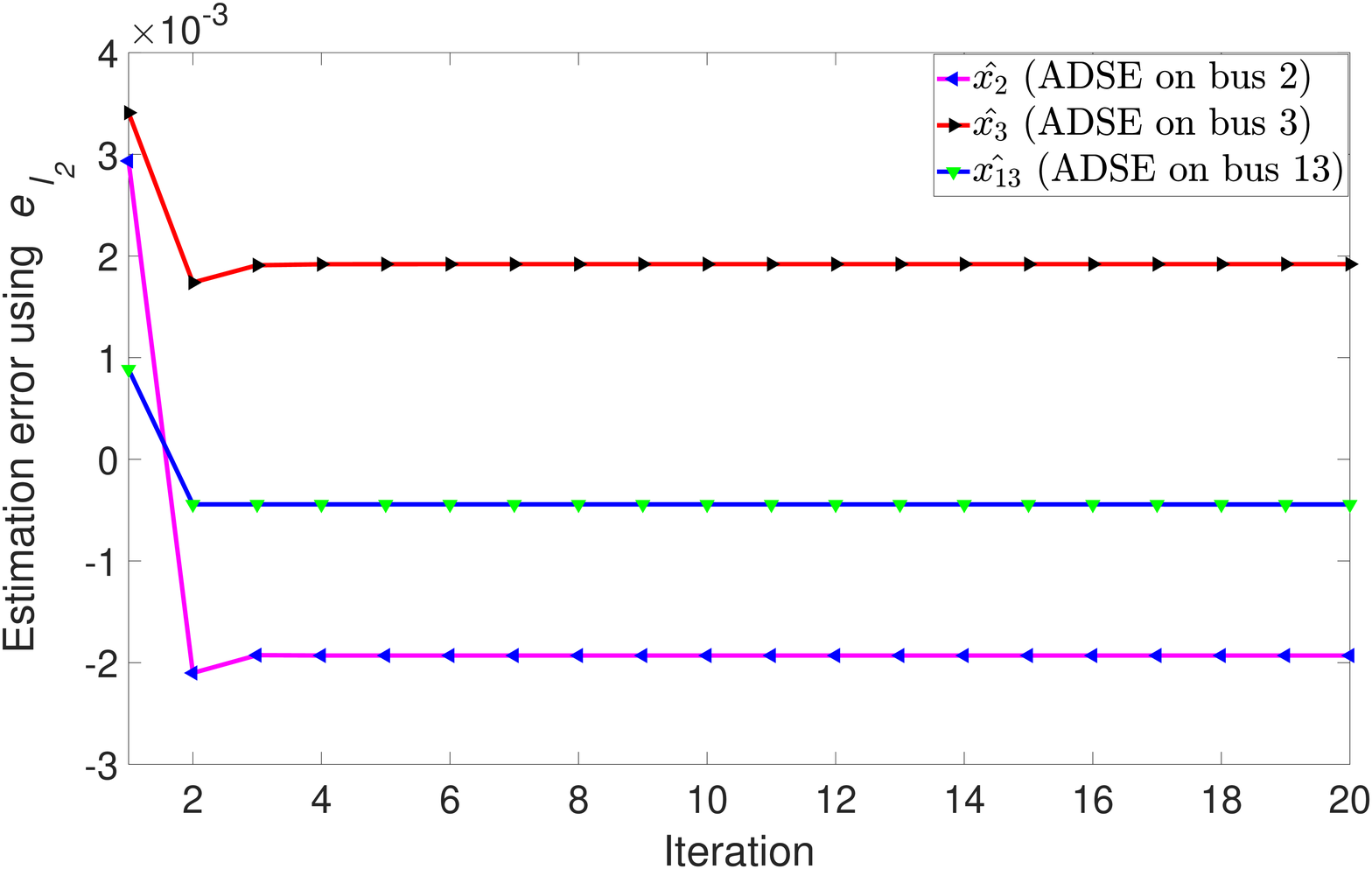}} \\
\subfloat[]{\includegraphics[width=1\linewidth]{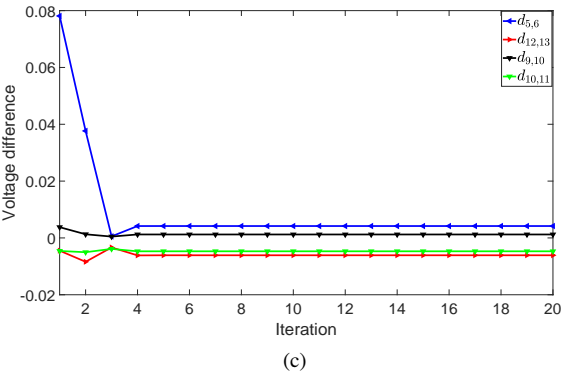}}
\caption{Estimation errors ($\sigma_w^2$ = $10^{-4}$) (a) Using $e_{l_2}$, (b) Using $e_{l_2}$ of selected buses, (c) Using $d_{\upsilon, \varphi}^i$ for the voltage difference.} \label{est2}
\end{figure}
The experiments of the developed framework are conducted using MATLAB R2019B. Measurement dataset is generated based on Fig. \ref{fig1} where relevant power system parameters including bus/branch data, admittance matrices, and the Jacobian matrices are extracted using MATPOWER \cite{zimmer2016matpwr}. First, estimation of system states is computed under normal circumstance using the iterative procedures of Algorithm \ref{admm} of the ADSE. Next, the performance evaluations of the proposed ADSE are compared to the centralised SE approach, which serves as a benchmark for power system SE tasks. The proposed ADSE is expected to have an approximated estimation accuracy to that of the centralized approach. Therefore, the estimation performances of the ADSE are benchmarked against the commonly known WLS-based \cite{abur2004power} \cite{liu2011false} centralised state estimator. Further, the ADSE is evaluated with respect to the proposed AG1 and AG2 considering the communication architecture given in Fig. \ref{comm}.
\begin{figure}[h!] \centering
\subfloat[]{\includegraphics[width=0.8\linewidth]{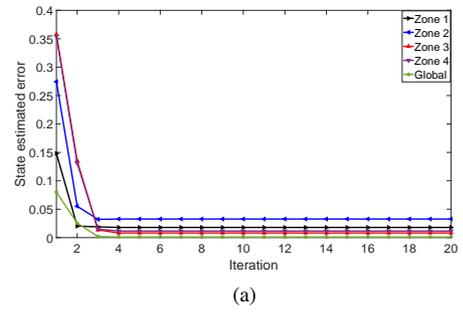}} \\
\subfloat[]{\includegraphics[width=0.8\linewidth]{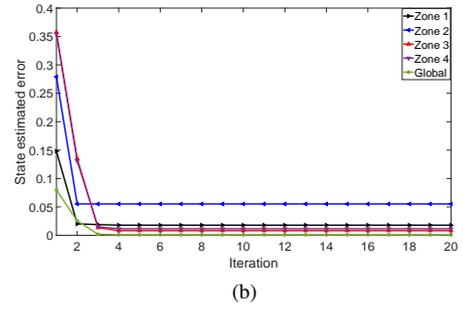}} \\
\subfloat[]{\includegraphics[width=0.8\linewidth]{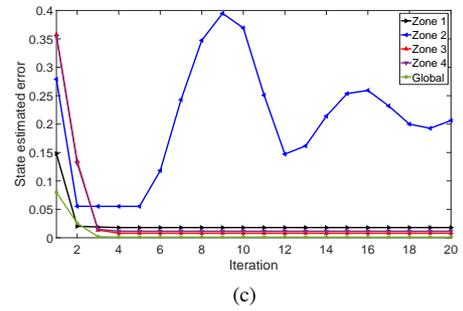}} \\
\subfloat[]{\includegraphics[width=0.8\linewidth]{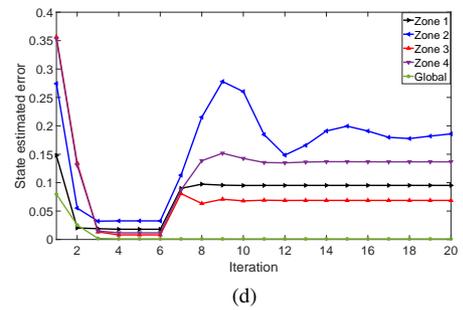}}
\caption{Comparison of local and global state estimated error curves of ADSE (a) under normal circumstances and when zones exchange boundary data (b) under AG1 (considering only availability attack), (c) under AG1 (considering availability and integrity attacks), and (d) under AG2.} \label{est3}
\end{figure}
\subsection{Results and Discussion}
The following error metrics are adopted under three different scenarios. First, state estimated error \begin{math}
(e_{ADSE}=\hat{\mathbf{x}}_\kappa^i - \mathbf{x}_\kappa^i)_{\forall \kappa, \forall i}
\end{math} between estimations of each local zone and ground truth states, as well as  estimations of each local zone and the centralised WLS state estimator is considered under normal circumstances. Similarly, the global estimation result is compared to the ground truth states and the centralised WLS state estimator. This evaluation metric helps us to assess the estimation performance of the ADSE against the WLS technique. Another metric is estimation error based on the $l_2$-norm where the estimation of each zone and the global level is compared to the corresponding ground truth system states, and to the centralised WLS state estimator. The $l_2$-norm based estimation error is given by
\begin{math}
e_{l_2} = \frac{||\hat{\mathbf{x}}_\kappa^i - \mathbf{x}_\kappa^i||_2}{||\mathbf{x}_\kappa^i||_2}\end{math} where $\hat{\mathbf{x}}_\kappa^i$ and $\mathbf{x}_\kappa^i$, respectively refer to the vectors of estimated and ground truth system states. Furthermore, to verify the impact of the incumbent cyberattacks against the distributed systems, state estimated errors of the local and global DSE under AG1 and AG2 are considered. Under this scenario, various levels of injection of attack magnitude are considered.

\subsubsection{Scenario I (under normal condition)} 
First, the estimation results of the ADSE are plotted in Fig. \ref{est1}, also compared to the WLS-based centralised SE and ground truth states. While Fig. \ref{est1}a is estimation results of the WLS-based centralised SE, Fig. \ref{est1}b is the global estimation result of the ADSE. Results of these two are also compared to the ground truth states in Fig. \ref{est1}c. Next, $e_{l_2}$ is shown in Fig \ref{est2}a, showing the errors between the global estimated values and ground truth states, and between the global estimated values and centralised WLS. Additionally, Fig. \ref{est2}b is estimation error using the $e_{l_2}$ of some selected buses (e.g. buses 2 from $Z1$, 3 from $Z2$, and 13 from $Z3$ are used). Finally, the deviation between estimation differences of bus indices to the centralised WLS is considered. This error metric is similar to the one adopted in \cite{xie2012fully}. The deviation between estimation differences between two different bus indices of $\upsilon$ and $\varphi$ is given by 
\begin{math}
d_{\upsilon, \varphi}^i = |(\hat{x}_\upsilon^i-\hat{x}_ \varphi^i)^{\text{WLS}}-(\hat{x}_\upsilon^i-\hat{x}_ \varphi^i)^{\text{ADSE}}|
\end{math}, where $|.|$ is the absolute value operator. This result is depicted in Fig. \ref{est2}c. 
Under Scenario I, the mean error in terms of the $e_{l_2}$ of the WLS-based SE and the ADSE is about 0.095\% and 0.11\%, respectively. Likewise, the estimation error in terms of MSE of the WLS-based SE and the ADSE is about 0.000093\% and 0.00016\% respectively. These estimation errors are acceptable for the EMS of the power system \cite{abur2004power}. 

The numerical results shown above demonstrate that, under normal operating conditions, the proposed ADSE method converges within acceptable estimation errors. It can also be shown that the estimation performances, including under measurement noises, are close to the results of the WLS estimator.

\subsubsection{Scenario II (Under AG1)}
To verify the impact of AG1, first, estimation errors of the local and global DSE under normal circumstances are considered, shown in Fig. \ref{est3}a. Then, the error between estimations obtained using (\ref{attAG11}) (i.e. after availability attack on the shared boundary data), and ground truth states of each local zone is shown in Fig. \ref{est3}b. Starting from iteration 2 (where the availability attack occurs) the estimation error of Fig. \ref{est3}b is higher compared to the estimation error under the normal circumstances of Fig. \ref{est3}a. For the integrity attack, let the adversary targets sensors connected to bus 4 of $Z_2$ (i.e. \{$M_4$, $M_{4-5}$, $M_{4-7}$, $M_{3-4}$\}). Note that although $Z_2$ is taken as an example, without loss of generality the attack
vector can be originated from other zones as well. Based on the targeted measurements, the attacker can cunningly select the injection vector $\mathbf{b}_\kappa$ to construct a non-zero attack vector \begin{math}
\mathbf{a}_\kappa=
    \mathbf{H}_\kappa\mathbf{b}_\kappa 
\end{math} as defined by (\ref{randm}). In this regard, let $\mathbf{b}_\kappa$ for bus 4 of $Z_2$ is $\alpha$ times the nominal voltage magnitude $b_0$, which is given by (\ref{trgt}).
\begin{equation}
 \mathbf{b_\text{2, 4}} \overset{\Delta}{=} [\underbrace{0,\overbrace{\alpha\times b_0}^\text{bus 4}, 0, 0}_\text{magnitude}, \underbrace{0,..,0}_\text{angle}]^T 
 \label{trgt}
\end{equation}
Then, the attack vector is obtained by replacing (\ref{trgt}) in \begin{math} \mathbf{a}_\kappa=
    \mathbf{H}_\kappa\mathbf{b}_\kappa 
\end{math}. Let $\alpha = -15\%$ for the data integrity attack of Scenario II and Scenario III although different values can be possible. The local unobservability attack considering the availability and integrity attacks is shown in Fig. \ref{est3}c. The estimation result under the AG1 of both the availability and integrity attacks shows much higher average estimation error (i.e. 34.74\% in terms of $e_{l_2}$). Accordingly, As there is no boundary communication among the neighbor zones, the local unobservability is limited within Zone 2.
\subsubsection{Scenario III (Under AG2)}
Now the impact of the false injection attack (\ref{trgt}) originated from zone $Z_2$ is evaluated under AG2. Under this scenario (shown in Fig. \ref{est3}d), the exchange of boundary data among the zones is considered as demonstrated in Section \ref{inteAttk}. Consequently, this attack affects both $Z_2$ and the rest of the zones. Under this scenario, the neighboring zones exchange boundary data and the local zones are compromised, resulting in estimation error ($e_{l_2}$) of 9.73, 21.51, 6.33, and 13.87\% respectively for $Z_1$, $Z_2$, $Z_3$, and $Z_4$.

\section{Conclusion} \label{concln}
This article has provided an in-depth analysis of the security threats of IIoT-based smart grid and highlighted serious security vulnerabilities. It was shown that a coordinated two-stage data availability and integrity attack can destabilise local zones of the distributed smart grid. Moreover, it was demonstrated how the data integrity attack can actually destabilise not only a particular zone but also the entire grid when the communication of the distributed system occurs properly. While the estimation error of the DSE is acceptable for smart grid EMS under normal operating conditions, it becomes unbounded (both locally and across the integrated grid) in the presence of orchestrated cyberattacks. Hence, the investigation of cyberattacks in the DSE in this article can help system designers foresee any potential interaction between the grid and adversaries. In consequence, it calls for solutions against the DSE vulnerabilities to improve the security and widespread adoption of IIoT-based smart grid ecosystem. 

 Future work recommendations can be taken into account, both from the perspective of the adversaries and system operators. As coordinated cyberattack against the DSE may have a wide range of security consequences such as energy theft, risk and reliability, secure operation and stability, electricity market and pricing economics, and so on, the study of threat models based on these impacts is an open research issue. Moreover, from the system operator/defence perspective, one potential future research can be if it is possible to develop a distributed attack detection using, for example, a game-theoretic approach leveraging the grid-adversary interaction. Further, after the attack identification, the operator may then need to mitigate compromised measurements (for example, leveraging pseudo-measurements) using data-driven approaches. Last but not least, to further deter the incumbent security challenges across the distributed smart grid, preventative security measure against the incumbent cyberattacks is another open research issue. 
\bibliographystyle{unsrt} 
\bibliography{Haftu_Bib_ElsarticleSC} 
\end{document}